\documentclass[runningheads]{llncs}
\usepackage[T1]{fontenc}
\usepackage{graphicx}
\usepackage{amsmath}
\usepackage{amssymb}
\usepackage{booktabs}
\usepackage{multirow}
\usepackage{amsmath}
\usepackage{array}
\usepackage{diagbox}
\usepackage{enumitem,amssymb}
\setlist[itemize,1]{label=\textbf{$\bullet$}}
\begin{document}
\title{Enhancing Medical Lesion Segmentation via Several Simple Modules}
%
%
\author{Ruichen Li\inst{1}\orcidID{0009-0001-0262-041X} \and
        Zhao Yao\inst{2}\orcidID{0009-0009-3609-7741}}
\authorrunning{R. Li and Z. Yao}
%
\institute{Hunan University, Changsha, China\\
\email{richardli@hnu.edu.cn}}
\maketitle              
\begin{abstract}
This work introduces three lightweight enhancement modules for medical lesion segmentation to improve clinical applicability, particularly under data-scarce and computationally constrained environments. Current mainstream methods struggle with the lack of high-quality annotations for rare pathologies, which often leads to severe morphological distortions during inference, despite achieving seemingly ideal DSC scores. Furthermore, the isolated development of individual segmentation tasks restricts cross-domain methodological innovation, underscoring the need for leveraging shared knowledge across different lesions within the same anatomical region. However, constrained access to historical data and annotations hinders efficient knowledge extraction and triggers catastrophic forgetting when adapting to new classes. To address these challenges, we first propose DT-Morph Combined Loss with continuous spatial pooling to optimize differentiable topological structures and boundary continuity. Second, we introduce 3D Group Prompt, which extracts domain-universal knowledge via curated volumetric image-mask alignment prompts together with text prompts built on SAT baseline. Finally, we replace conventional output layers with Many-to-One Conditional Segmentation Head equipped with synonym-resolved routing, which brings negligible computational cost and enables fast incremental learning for new categories without sacrificing performance on old tasks. Ablation studies on abdominal and extra-abdominal datasets covering both organ and lesion segmentation verify the efficacy of our modules.

\keywords{Tumor Segmentation  \and Vision-Language Model  \and Prompt Learning \and CT Morphology.}
\end{abstract}
\section{Introduction}

Medical image segmentation underpins clinical automated diagnosis, yet precisely segmenting tiny, morphologically irregular lesions remains a critical challenge. Deep learning methods have shown significant success here, but they require complete access to large-scale, high-quality annotated data during training. This requirement rarely holds in clinical practice, where annotations for rare pathologies are often scarce and difficult to obtain.

Segmentation-loss research has evolved from pixel-wise region optimization to shape-constrained regularization, yet classic objectives still focus on global overlap and voxel-wise accuracy [1], while most morphological priors are non-differentiable and 2D-limited, producing fragmented predictions and hollow artifacts for small, topologically complex lesions [3]. To address this, we propose the DT-Morph Combined Loss, which realizes differentiable 3D morphological closing via continuous spatial pooling and confines distance-transform penalization to fine-grained edge voxels with dynamic boundary masks [4]. Combined with Otsu auto-binarization and multi-scale geometric regularization, it eliminates discretization artifacts and sustains topological continuity [2].

Beyond fine-grained accuracy, robust cross-domain generalization is another core thread. Recent prompt-driven paradigms improve transfer performance, yet mainstream slice-wise 2D frameworks fail to model inter-slice spatial continuity of volumetric imaging [6], causing severe out-of-bounds false positives and degraded 3D generalization on heterogeneous backgrounds [7]. To resolve this, we propose GroupPrompt3D, a lightweight generalized segmentation framework. A depth context module aggregates inter-slice continuity with negligible overhead [8], while a lightweight 3D ASPP with depthwise separable dilated convolutions cuts multi-scale 3D computation by over 90\% [9]. Coupled with a Cross-Transformer that dynamically generates condition-aware prompt kernels, it outperforms the 1000-epoch baseline with only 200 epochs, achieving superior zero-shot and few-shot generalization [5].

Additionally, scalable universal segmentation with semantic consistency is another dominant direction. Existing text-guided frameworks mostly adopt one-to-one label-to-head mapping [12], overlooking widespread synonym ambiguity and inconsistent label taxonomies across clinical protocols [11], triggering drastic performance drops under lexical perturbation and catastrophic forgetting in incremental learning. To tackle this, we develop the Many-to-One Conditional Segmentation Head, equipped with a medical synonym-to-canonical mapping layer and multi-source embedding-guided parameter generation. It unifies semantically inconsistent labels via global parameter sharing across a lightweight Meta-MLP, staying stable under synonym substitution and enabling flexible category extension without previous-knowledge erosion [10].

We would like to note that these modules are complementary and integrate into existing frameworks with negligible overhead. Extensive ablation experiments demonstrate our state-of-the-art performance in fine-grained lesion segmentation, 3D cross-domain generalization, and scalable universal segmentation.

\section{Related Work}
\subsection{Universal and Text‑Guided Medical Image Segmentation.}
Recent medical foundation models increasingly favor unified architectures that handle diverse modalities and anatomical targets. In text-guided segmentation, SAT [13] segments 497 anatomical classes via free-form text prompts with a knowledge-enhanced design, alongside interactive prompt-driven paradigms pioneered by SAM [15]. However, processing all voxels uniformly without explicit topological or geometric priors, such models often underperform on small, irregular lesions prone to fragmentation and discretisation artifacts. This limitation, shared by prompt-guided variants like MedSAM-3 [14], persists despite their flexibility with open-vocabulary descriptions, as they still lack the fine-grained shape regularisation and precise 3D spatial coherence required for challenging anatomies.

\subsection{Tumor Segmentation and Prompt Learning.}
Tumour segmentation has been advanced by prompt-learning frameworks like UniCDSeg [16], that adapt to diverse tasks with minimal re-training via cross-domain dynamic kernels, yet their inherent 2D design lacks the volumetric continuity crucial for 3D diagnosis. To capture 3D spatial context, volumetric backbones have evolved rapidly: UNETR [18] offers long-range modeling but suffers from quadratic complexity; SwinUNETR-V2 [19] mitigates this via windowed attention and residual convolutions; MedNeXt [20] modernizes 3D CNNs with deep kernel scaling for robust boundary extraction; SegMamba [17] achieves linear complexity via State Space Models for full-volume modeling. Extending low-complexity 3D backbones like SegMamba into dynamic prompt-learning frameworks remains an open challenge.

\subsection{Vision–Language Alignment and Synonym Resolution.}
To overcome label taxonomy inconsistencies and partial annotations, vision–language alignment has emerged as a powerful solution. The CLIP-Driven Universal Model [22] replaces fixed output channels with a language-driven parameter generator that utilises pre-trained textual embeddings from CLIP [21], enabling adaptive segmentation across diverse datasets; variants like Med-VLM [23] further incorporate textual descriptors to refine masks. However, these methods enforce a rigid one-to-one mapping between textual labels and conditional heads, which may cause gradient scattering and inconsistent predictions across datasets. Recent efforts like MHG-Adapter [24] mitigate this via hypergraph structures, yet a systematic solution for synonym resolution remains elusive.

\section{Methodology}

\subsection{DT-Morph Combined Loss.}
Traditional region-based objectives, like the Sample-Wise Binary Dice Loss, optimize voxel-wise distributions independently—disregarding underlying geometry and boundary smoothness. This limitation causes fractured predictions and erratic contours, which severely amplifies task interference during task vector composition. To mitigate this, we introduce a Differentiable Distance-Transform Morphological Loss ($\boldsymbol{L}_{\text{DTMorph}}$). By embedding topological and boundary priors directly into the backpropagation pipeline, $\boldsymbol{L}_{\text{DTMorph}}$ acts as a potent structural regularizer to harmonize task vectors before merging.
\subsubsection{Differentiable Topological Closing.}

We simulate the morphological closing operation—traditionally non-differentiable—by leveraging continuous spatial pooling operations. The soft closing operator $\boldsymbol{\mathcal{C}}_k(\cdot)$ applied to the prediction map $\boldsymbol{\hat{Y}} = \boldsymbol{p}_{\theta}(\boldsymbol{y}|\boldsymbol{x})$ utilizes sequential 3D maximum pooling and its inverse with a specified structural kernel size. The morphological loss $\boldsymbol{L}_{\text{Morph}}$ penalizes structural and topological discrepancies by evaluating the mean squared error between the closed representation of the predicted probability maps and that of the ground-truth target masks. This operation selectively fills small internal perforations and bridges disconnected segments, forcing the network to respect global anatomical connectivity rather than isolated voxel accuracy.

\begin{figure}
\centering
\includegraphics[width=\textwidth]{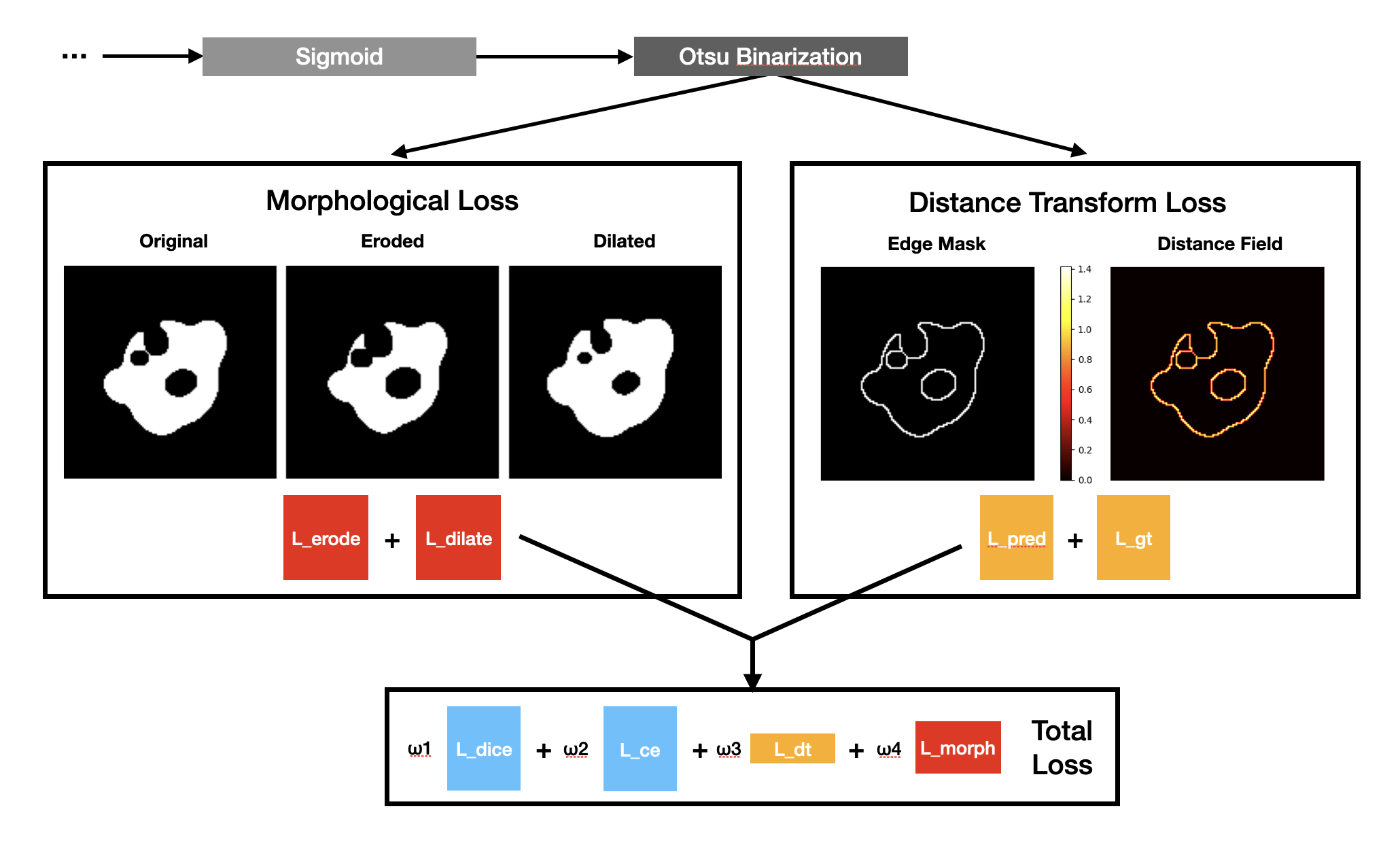}
\caption{Overview of DT-MorphLoss. The overall pipeline proceeds as follows: 1) a Sigmoid activation and Otsu binarization convert the prediction map into a binary mask; 2) the mask is fed in parallel into two branches, where the morphological-loss branch applies erosion and dilation to compute $L_{\text{morph}}$, while the distance-transform branch extracts an edge mask to produce a distance field for $L_{\text{dt}}$.; 3) the two terms are weighted and combined with the standard Dice loss ($L_{\text{dice}}$) and cross-entropy loss ($L_{\text{ce}}$) to obtain the total loss.} \label{fig1}

\end{figure}

\subsubsection{Soft Boundary Alignment.}
To further refine high-frequency geometric details, we implement a soft edge extractor $\boldsymbol{\mathcal{E}}(\cdot)$ to capture boundary alignment. Soft boundary maps are obtained by computing the pixel-wise difference between the dilated and eroded volumes of the segmentations. The edge alignment objective is subsequently defined as the mean squared error between the extracted boundary maps of the prediction and the ground truth. The unified geometric penalty combines boundary alignment with a local neighborhood smoothness constraint $\boldsymbol{L}_{\text{Smooth}}$ ($\boldsymbol{L}_{\text{DT}} = 0.8 \boldsymbol{L}_{\text{Edge}} + 0.2 \boldsymbol{L}_{\text{Smooth}}$).

\subsubsection{Multi-Stage Optimization Strategy.}

To guarantee stable convergence, the morphological and geometric constraints are dynamically introduced following a warm-up period $\boldsymbol{T}$ :

\begin{equation}
\boldsymbol{L}_{\text{Total}}(\boldsymbol{\theta}) = \mathbb{I}(\boldsymbol{t} \le \boldsymbol{T}) \cdot \boldsymbol{L}_{\text{Base}}(\boldsymbol{\theta}) + \mathbb{I}(\boldsymbol{t} > \boldsymbol{T}) \cdot \Big( 0.8 \boldsymbol{L}_{\text{Base}}(\boldsymbol{\theta}) + \boldsymbol{\lambda}_1 \boldsymbol{L}_{\text{DT}}(\boldsymbol{\theta}) + \boldsymbol{\lambda}_2 \boldsymbol{L}_{\text{Morph}}(\boldsymbol{\theta}) \Big)
\end{equation}

where $\boldsymbol{L}_{\text{Base}}(\boldsymbol{\theta}) = 0.5 \boldsymbol{L}_{\text{Dice}} + 0.5 \boldsymbol{L}_{\text{BCE}}$. By transiting from region-only constraints to a joint morphological-geometric formulation, individual models achieve topologically complete task vectors, substantially mitigating parameter-level interference during the subsequent merging operation.

\subsection{3D Group Prompt.}
Based on the framework of Spider-UniCDSeg, our GroupPrompt3D implementation introduces several key methodological advancements to transition from 2D slice-based processing to fully 3D volumetric awareness. We formalize this universal, prompt-driven architecture as a mapping over a 3D query volume $\boldsymbol{x}_q$ to predict voxel-wise class distributions $\boldsymbol{p}_{\theta}(\boldsymbol{y}|\boldsymbol{x}_q) $. The network behavior is conditionally modulated by an explicit support prompt pair $\boldsymbol{\mathcal{S}} = \{\boldsymbol{x}_s, \boldsymbol{m}_s\}$, where $\boldsymbol{x}_s$ denotes the reference support volume and $\boldsymbol{m}_s$ denotes its corresponding binary morphological mask.

\subsubsection{Volumetric Reshaping and Multi-Scale Feature Extraction.}

To bridge the optimization gap between high-capacity 2D representations and 3D geometric spaces, any input volume $\boldsymbol{x}$ is flattened along the depth axis into a 2D slice sequence $\boldsymbol{\tilde{x}} \in \mathbb{R}^{(\boldsymbol{B} \cdot \boldsymbol{D}) \times \boldsymbol{C}_{in} \times \boldsymbol{H} \times \boldsymbol{W}}$ for processing through a frozen Swin-Transformer Large backbone. The multi-scale 2D feature hierarchies are immediately restored into 3D tensors and projection-aligned into intermediate 3D features:
\begin{equation}
\boldsymbol{\mathcal{T}}_i = \boldsymbol{\psi}_i\left( \boldsymbol{\mathcal{R}}\left( \textbf{Backbone}_i(\boldsymbol{\tilde{x}}) \right) \right) \in \mathbb{R}^{\boldsymbol{C}_i \times \boldsymbol{D} \times \boldsymbol{h}_i \times \boldsymbol{w}_i}, \quad \boldsymbol{i} \in \{2, 3, 4, 5\}
\end{equation}

where $\boldsymbol{\mathcal{R}}(\cdot)$ denotes the spatial dimensional restoration operator, and $\boldsymbol{\psi}_i(\cdot)$ denotes channel capacities.

\subsubsection{Feature Aggregation.}

To restore continuous morphological boundaries across variable lesion scales without losing spatial resolution, the deepest semantic layer $\boldsymbol{\mathcal{T}}_5$ is regularized via a 3D Atrous Spatial Pyramid Pooling (\textit{ASPP3D}) layer. It aggregates multi‑scale context via parallel branches: a standard convolution block, multiple 3D dilated convolutions with varied dilation rates, and a 3D adaptive average pooling layer with trilinear upsampling. Multi‑branch outputs are channel‑concatenated for fusion. Subsequently, a top-down Gated Feature Pyramid Network (Gated \textit{FPN}) recursively fuses upsampled deep semantic features with shallow spatial details. Through gated convolution blocks and trilinear interpolation, this decoder generates a single high‑resolution feature map $\boldsymbol{\mathcal{D}}_2$ enriched with both global semantics and local boundary details.

\begin{figure}
\centering
\includegraphics[width=12cm]{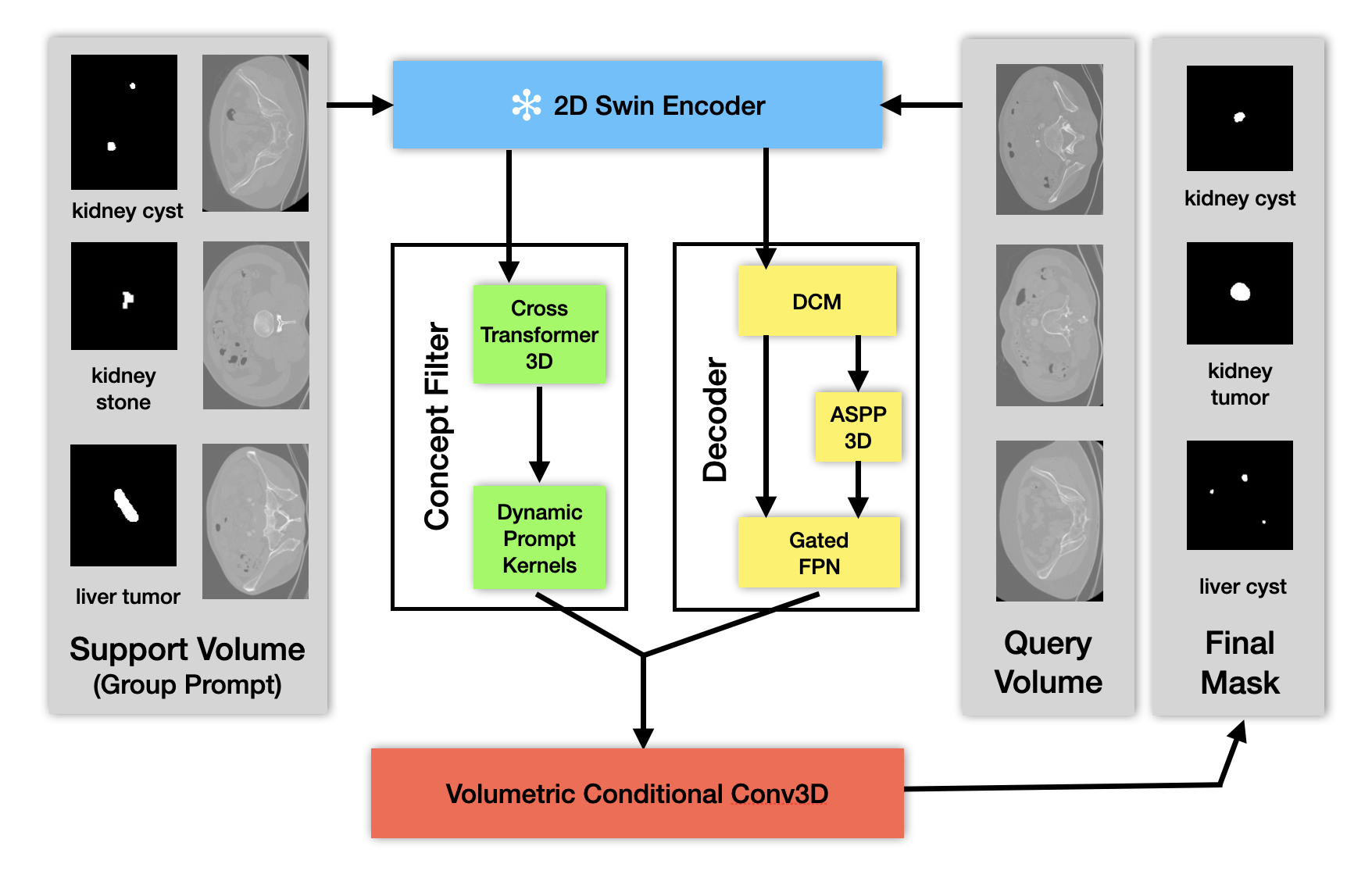}
\caption{Overview of GroupPrompt3D. A 2D Swin encoder extracts features from both a support volume and a query volume. The support path processes features through a concept filter, while the query path extracts multi-scale representations via a decoder. Crucially, since the target labels in the support and query volumes do not perfectly match during training, a Conv3D block is required to map the conditional prompt knowledge and predict the final mask.}
\end{figure}

\subsubsection{Dual-Prototype Cross-Attention and Prompt Execution.}

To isolate background clutters in 3D volumes, we refine the conditional filter logic of UniCDSeg by constructing dual-prototypes (foreground and background). Let $\boldsymbol{j}$ index the spatial-depth coordinate grid $(\boldsymbol{h}, \boldsymbol{w}, \boldsymbol{d})$. The target-specific prototype vectors $\boldsymbol{\mathcal{P}}_{fore}, \boldsymbol{\mathcal{P}}_{back} \in \mathbb{R}^{1 \times \boldsymbol{C}}$ are extracted via masked pooling:
\begin{equation}
\boldsymbol{\mathcal{P}}_{fore} = \frac{\boldsymbol{\sum}_{j} \boldsymbol{\mathcal{T}}_5(j) \cdot \boldsymbol{m}_s(j)}{\boldsymbol{\sum}_{j} \boldsymbol{m}_s(j)}, \quad \boldsymbol{\mathcal{P}}_{back} = \frac{\boldsymbol{\sum}_{j} \boldsymbol{\mathcal{T}}_5(j) \cdot (1 - \boldsymbol{m}_s(j))}{\boldsymbol{\sum}_{j} (1 - \boldsymbol{m}_s(j))}
\end{equation}

These prototypes act as queries within a CrossTransformer3D module to interact with the global context of the support volume. Finally, a linear projection maps $\boldsymbol{\mathcal{P}}^*$ into a dynamic weight matrix and a bias vector. The final query volume segmentation is executed via a 3D convolutional layer that applies these dynamically synthesized weights and biases over the unified feature map of the query volume, achieving an environment-adaptive inference tailored to the reference support context.

\begin{table}
\centering
\caption{Improvements achieved by GroupPrompt3D.}\label{tab1}
\begin{tabular}{|l|l|l|}
\hline
Feature &  GroupPrompt3D & Spider-UniCDSeg (Base)\\
\hline
Spatial Dimension & {\bfseries Full 3D volumetric} & 2D or pseudo-3D\\
Inter-Slice Modeling & {\bfseries Explicit} & Implicit\\
Feature Aggregation & {\bfseries 3D separable ASPP} & Standard ASPP\\
Guidance Logic & {\bfseries 3D cross-attention kernels} & Feature conditioning\\
Input Flexibility & {\bfseries Dynamic crop sizes} & Fixed sizes\\
\hline
\end{tabular}
\end{table}

\subsection{Many-to-One Conditional Segmentation Head.}

Aiming to mitigate segmentation performance degradation under new or semantically ambiguous labels, we introduce an enhanced Conditional Segmentation Head (\textit{CSH}) based on CLIP-Driven-Universal-Model (CLIP-UM). By decoupling spatial localization from semantic abstraction, this network adaptively segments diverse anatomical and pathological structures without increasing base parameters. The framework comprises a multi‑scale encoder‑decoder backbone and dynamic conditional segmentation heads. The backbone extracts multi‑scale latent embeddings $\boldsymbol{\mathcal{F}}_{\text{lat}}$ and per-pixel embeddings $\boldsymbol{\mathcal{F}}_{\text{pix}}$, where the finest spatial embedding is denoted as $\boldsymbol{U}_i$. Meanwhile, each textual label $\boldsymbol{y}_n$ passes through a semantic dictionary to trigger a canonical parameter generator, which dynamically synthesizes 3D convolutional kernels to decode mask $\boldsymbol{m}_n$ from $\boldsymbol{U}_i$.

\subsubsection{Dynamic Parameter Generation.}

The CSH leverages a hypernetwork paradigm to generate content-adaptive parameters on the fly. The head constructs an aggregated descriptor vector $\boldsymbol{x}_{b,q}$ for each batch via the concatenation of three distinct latent representations: the explicit instance-level query embedding $\boldsymbol{Q}_{b,q}$, the task-level group prompt embedding $\boldsymbol{P}_b$ and re-projecting the group features via a linear layer $\boldsymbol{\phi}_{\text{group}}$, and the visual gap embedding $\boldsymbol{V}_b$ extracted by spatial average pooling of the deepest backbone bottleneck feature $\boldsymbol{B}_{\text{layer}}$ and projecting it via a linear mapping $\boldsymbol{\phi}_{\text{visual}}$. The aggregated vector $\boldsymbol{x}_{b,q}$ is then passed into a multi-layer perceptron (\textit{MLP}) mapping function $\boldsymbol{H}_k$ activated by GELU non-linearities, where $\boldsymbol{k}$ denotes the canonical sub-head index, to yield the parameter vector $\boldsymbol{\Theta}_{b,q}$. The dimension of this bottleneck parameter space is defined as $\textbf{Dim}(\boldsymbol{\Theta}_{b,q}) = (\boldsymbol{M} \times \boldsymbol{C}_{\boldsymbol{u}_i}) + \boldsymbol{M}^2 + 3\boldsymbol{M}$ + 1, where $\boldsymbol{M}$ denotes the intermediate channel width, and $\boldsymbol{C}_{u_i}$ denotes the per-pixel feature channels of $\boldsymbol{U}_i$.

\subsubsection{Conditional Dynamic Convolution.}

The generated parameter vector $\boldsymbol{\Theta}_{b,q}$ is unrolled into a sequenced set of three-dimensional convolutional weights and biases, denoted as $\boldsymbol{W}=\{\boldsymbol{W}_1, \boldsymbol{b}_1, \boldsymbol{W}_2, \boldsymbol{b}_2, \boldsymbol{W}_3, \boldsymbol{b}_3\}$. Spatial mask prediction is executed sequentially through grouped 3D convolutions, denoted as $\boldsymbol{\circledast}$, using the batch size $\boldsymbol{B}$ as the number of parallel groups to map the continuous per-pixel embedding $\boldsymbol{U}_i$. Specifically, the intermediate feature map $\boldsymbol{Z}_1$ is generated by applying the weight $\boldsymbol{W}_1$ and bias $\boldsymbol{b}_1$ to $\boldsymbol{U}_i$ followed by a GELU activation. The second feature map $\boldsymbol{Z}_2$ is similarly computed by applying $\boldsymbol{W}_2$ and $\boldsymbol{b}_2$ to $\boldsymbol{Z}_1$ with GELU activation. The final generated logit slice $\boldsymbol{\hat{M}}_q$ for the $\boldsymbol{q}$-th query is obtained by convolving $\boldsymbol{Z}_2$ with the weight $\boldsymbol{W}_3$ and adding the bias $\boldsymbol{b}_3$.

\begin{figure}
\centering
\rotatebox{-90}{\includegraphics[width=10cm]{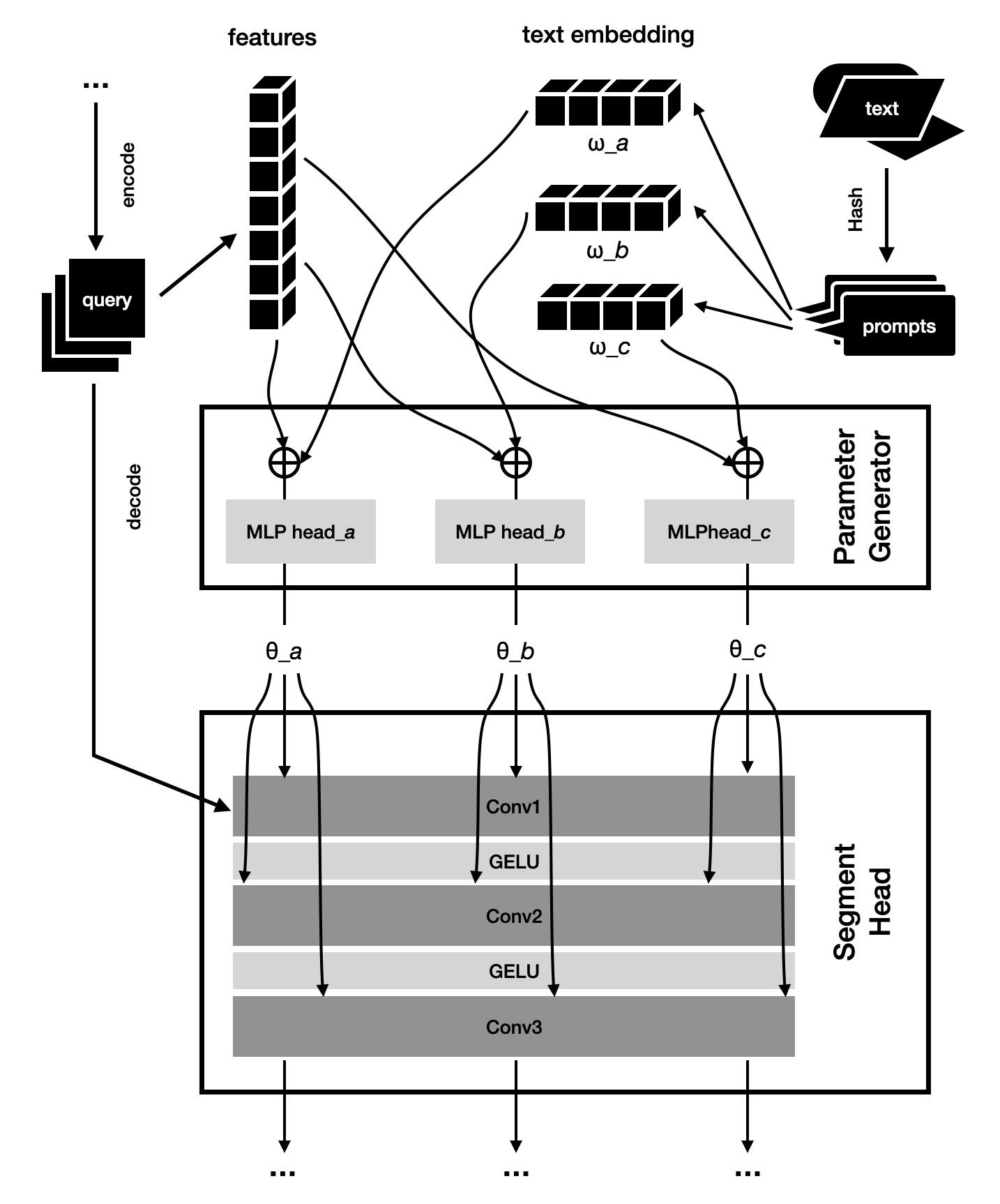}}
\caption{Overview of Many-to-one-CSH. Visual queries are encoded into multi-level features, while the text branch uses a Hash function to map various input texts into corresponding targets within a vocabulary registry and generate specific prompts. These prompts are transformed into distinctive text embeddings $\boldsymbol{\omega_i}$, which are sequentially combined with the visual features and fed into a parameter generator composed of multiple specialized MLP heads. This module dynamically generates prompt-conditioned weight parameters $\boldsymbol{\theta_i}$ that directly modulate the layers within the segment head, allowing the decoded query features to be processed into final masks.}

\end{figure}

\subsubsection{Synonym-Resolved Strategy.}

We implement a Many-to-One Semantic Routing protocol. Given a global medical nomenclature dictionary $\mathcal{\boldsymbol{V}}$ that maps a single canonical category $\boldsymbol{c}$ to an array of synonyms $\boldsymbol{S}_c$, an inverse resolution mapping $\boldsymbol{f}_{\text{res}}$ maps input synonyms back to their canonical forms. For each label token $\boldsymbol{y}$, we extract the first 20 characters of the SHA‑1 hash of its canonical token as a fixed‑length signature, which indexes ModuleDict to route inputs to the corresponding parameter generator $\boldsymbol{H}_k$. During training, a single-head multi-scale deep supervision mechanism is executed. The high-resolution mask logits $\boldsymbol{\hat{M}}$ generated from $\boldsymbol{U}_i$ are trilinearly interpolated to align with the spatial dimensions of lower-level pixel embedding layers, yielding the multi-scale predicted masks $\boldsymbol{\hat{M}}^{(l)}$. The overall objective function $\boldsymbol{\mathcal{L}}_{\text{total}}$ sums weighted compound segmentation losses between $\boldsymbol{\hat{M}}^{(l)}$ and the corresponding targets $\boldsymbol{Y}^{(l)}$ across all downsampled scales $\boldsymbol{l}$, each weighted by a deep supervision scale coefficient $\boldsymbol{\lambda}_l$.

\begin{table}
\centering
\caption{Improvements achieved by Many-to-one-CSH.}\label{tab1}
\begin{tabular}{|l|l|l|}
\hline
Feature &  Many-to-One-CSH & CLIP-Driven (Base)\\
\hline
Label Routing & {\bfseries Many to one} & One to one\\
Lexical Robustness & {\bfseries Pathology-centric} & Syntactic\\
Embedding Alignment & {\bfseries Multi source} & Dual source\\
Weight Projection & {\bfseries Deep non-linear} & Linear\\
Scalability & {\bfseries Dynamic} & Static\\
\hline
\end{tabular}
\end{table}

\section{Experiments}

\subsection{Datasets.}

We conduct experiments on 4 publicly available CT datasets, covering both tumor‑specific and multi‑organ general segmentation tasks:

\begin{itemize}
\item \textbf{MSWAL}: An abdominal tumor dataset containing 7 labes, serving as the primary benchmark for training and evaluation [25].
\item \textbf{AbdomenAtlas 3.0}: A large‑scale abdominal atlas, 4 tumor labels of which (gall bladder, kidney lesion, liver lesion, pancreatic lesion) are selected [26].
\item \textbf{AMOS22}: A large‑scale abdominal multi‑organ benchmark, comprises 16 anatomical labels, offering a rich source for evaluating generalisation [27].
\item \textbf{FLARE22}: Another multi‑organ dataset focused on general abdominal segmentation, comprises 15 anatomical labels [28].
\end{itemize}

\subsection{Results of DT-MorphLoss.}

Visual segmentation reveals the nonlinear evolution of geometric boundary constraints controlled by our loss function with different weight $\omega$. At $\omega=0.001$, weak boundary regularization causes severe over-segmentation, false positives, jagged edges and expanded contours. When $\omega$ rises to 0.05–0.15, dominant boundary penalties produce topological tearing: small targets become hollow rings, and tiny uncertain lesions are prematurely removed. The best balance occurs at $\omega=0.5$. Here boundary constraints coordinate well with Dice and CE regional losses, eliminating hollow defects, recovering truncated small lesions. By contrast, an extreme $\omega=2$ leads to structural breakdown. Excessive morphological erosion overrides regional loss, causing massive under-segmentation.

\begin{figure}
\centering
\includegraphics[width=12cm]{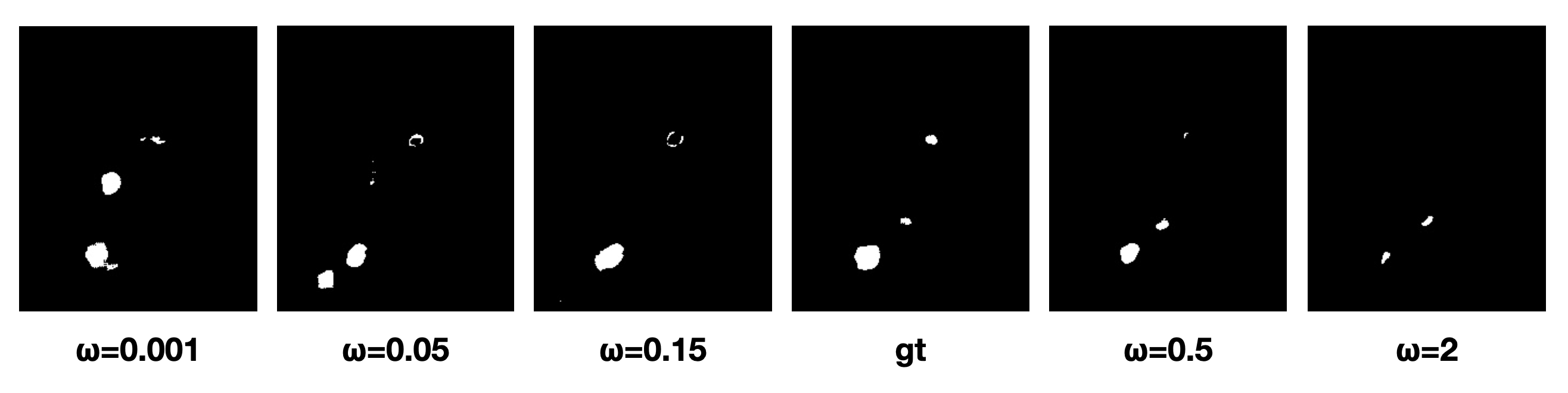}
\caption{Segmentation predictions versus the weight of DT-Morph Combined Loss for the identical liver tumor slice. The results reveal the nonlinear morphological evolution of the prediction, demonstrating the transition from over-segmentation to under-segmentation as $\omega$ increases.}

\end{figure}

We carried out hyperparameter tuning, confirming that scaling the loss weight toward $\omega=0.45$ consistently yields superior overall accuracy compared to unconstrained training. DCE quantitative results confirm the visual trends and further clarify cross-domain generalization and multi-task training behavior. To avoid excessive computation in long training cycles, we schedule dt-morph loss: the combined loss is disabled in the early learning stage, activated after 10,000 warmup epochs, and then dt loss is scaled down or shut down after 20,000 epochs, keeping only morph loss. This design avoids the heavy 3D computation cost of dt loss, which brings limited performance gains compared with lightweight morph loss. On the MSWAL and AbAtlas3 test sets, our model consistently outperforms the baseline as well as state-of-the-art architectures, proving that the 3D geometric prior acts as a general anatomical constraint. Comparison between 400 and 1000 training epochs demonstrates that extended training enables optimal boundary regularization to fully converge, while revealing divergent convergence across lesion types. Ambiguous lesions such as pancreatic cancer and liver cysts show late improvement in DCE (\textbf{0.290}, epoch=1000 / \textbf{0.262}, epoch=400). In contrast, liver tumors gain boundary advantages early and peak at 400 epochs (\textbf{0.266}, $\omega=0.45$). Moreover, small high-contrast calcifications such as kidney stones work best under weak constraints (\textbf{0.460}, $\omega=0.05$) and worsen under strong erosion (\textbf{0.346}, $\omega=0.15$).

\begin{table}[htbp]
\centering
\setlength{\heavyrulewidth}{1.2pt}
\setlength{\lightrulewidth}{0.6pt}
\setlength{\tabcolsep}{10pt}
\large
\caption{Ablation study results of DT-MorphLoss.}
\label{tab1}
\resizebox{0.9\textwidth}{!}{%
\begin{tabular}{l c c c | c c}
\toprule
\textbf{Dataset} & \textbf{Loss Weight} & \textbf{Epochs} & \textbf{Ours} & \textbf{Others} & \textbf{Dice} \\
\midrule
\multirow{5}{*}{\shortstack{MSWAL/\\AbAtlas3}} 
 & 0 (baseline)  & 1000 & 0.399 / 0.336 & UNETR        & 0.332 / 0.264 \\
 & 0.05 & 400  & 0.387 / 0.298 & MedNeXt      & 0.351 / 0.281 \\
 & 0.15 & 400  & 0.346 / 0.274 & SAT (baseline)      & 0.399 / 0.336 \\
 & 0.45 & 400  & 0.419 / 0.348 & SwinUNETR-V2 & 0.416 / 0.322 \\
 & 0.45 & 1000 & \textbf{0.460} / 0.366 & SegMamba & 0.457 / \textbf{0.368} \\
\bottomrule
\end{tabular}%
}
\end{table}

\subsection{Results of GroupPrompt3D.}

Slice-wise prediction results on multiple test cases verify that the proposed Group Prompt 3D module ($prompts = 60$) outperforms the baseline ($prompts = 0$) in spatial localization and prediction accuracy compared with ground truth. The core strength of this module is its powerful ability to suppress distant large-scale false positives. In Case15 (liver tumor) and Case14 (liver tumor), the baseline generates severe out-of-bounds false predictions covering unrelated slices, whereas the prompt-based model removes such isolated false outputs and restricts predictions strictly within target regions. Besides, the module improves boundary alignment and effectively reduces boundary overflow seen in the baseline on Case11 (liver cyst/tumor) and Case15 (kidney cyst). Nevertheless, the model still retains some drawbacks from the baseline network, including missed lesions in difficult regions (kidney tumor in Case14 and Case15) and local under-segmentation (kidney tumor in Case12). This reveals that prompt performance is largely constrained by the baseline network’s basic feature extraction ability.

\begin{figure}
\centering
\includegraphics[width=12cm]{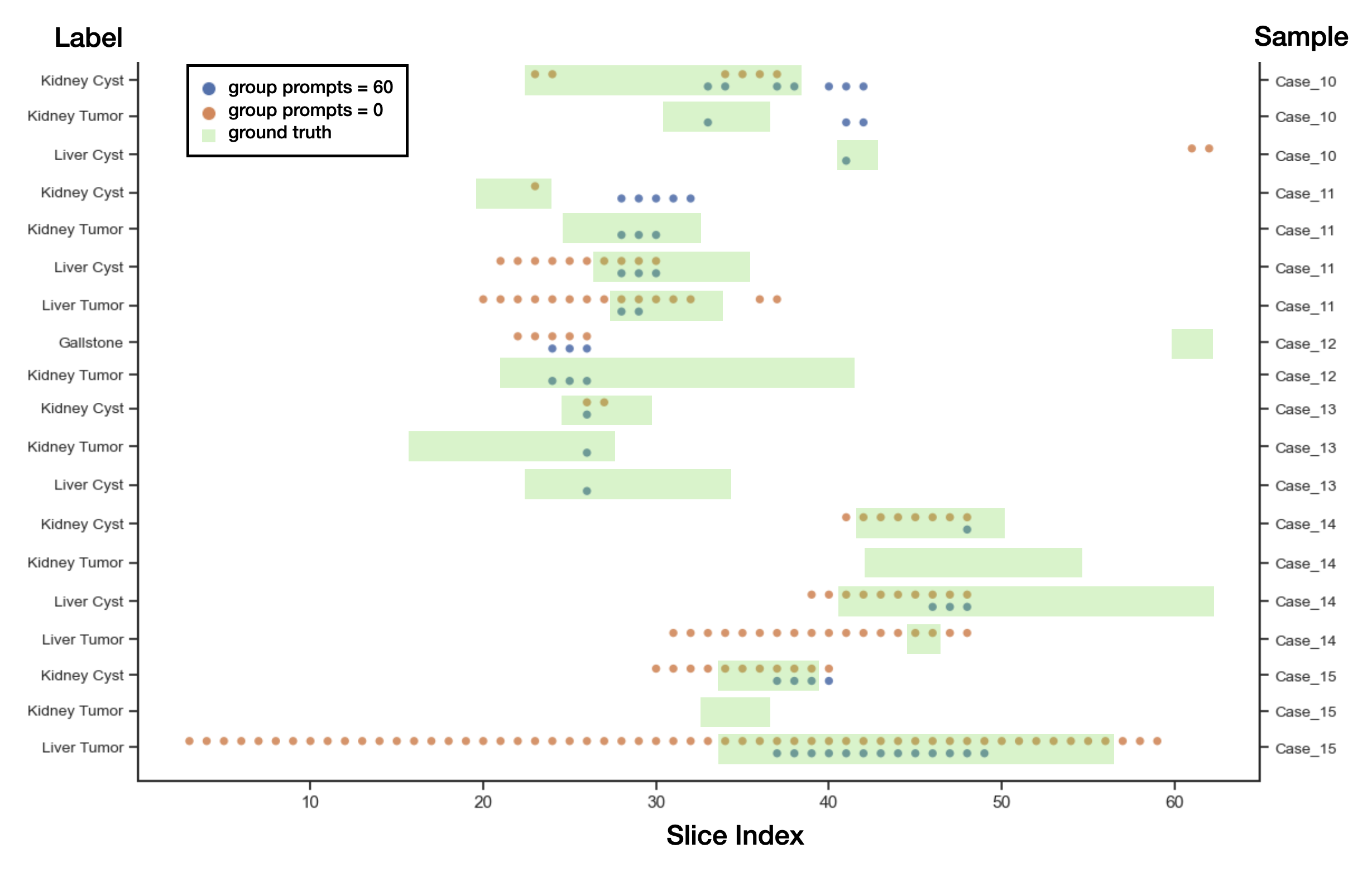}
\caption{Unified axial-space mapping of the discrete slice distributions across 6 test cases with 5 lesion categories. The horizontal axis indicates the actual slice index, while the dual-axis aligns the lesion labels and the case identifiers.}
\end{figure}

\begin{table}[htbp]
\centering
\setlength{\heavyrulewidth}{1.2pt}
\setlength{\lightrulewidth}{0.6pt}
\setlength{\tabcolsep}{10pt}
\large
\caption{Ablation study results of GroupPrompt3D.}\label{tab1}
\resizebox{0.7\textwidth}{!}{%
\begin{tabular}{l c c c c c c c}
\toprule
\textbf{Support Set} & \textbf{Prompts} & \textbf{Epochs} & \textbf{Dice}  & \textbf{Total Time}  \\
\midrule
N/A (baseline)  & 0 & 1000 & 0.433 & 25h \\
\midrule
\multirow{4}{*}{MSWAL}  & 15 & 200 & 0.394 & 11h \\
                        & 30 & 200 & 0.422 & 14h \\
                        & 60 & 200 & 0.444 & 21h \\
                        & 60 & 1000 & 0.484 & 104h \\
\midrule
\multirow{3}{*}{AMOS22}  & 15 & 200 & 0.387 & 12h \\
                         & 60 & 200 & 0.432 & 23h \\
                         & 60 & 1000 & 0.494 & 114h \\
\bottomrule
\end{tabular}%
}
\end{table}

Quantitative ablation experiments analyze how support set selection, prompt quantity and training epochs affect Dice metrics and computation cost. We adopted the K-means algorithm to sample support sets of different sizes from the original dataset, while maintaining balanced label distribution across all sampled cases. With training fixed at 200 epochs, increasing prompts from 15 to 60 continuously lifts Dice scores on both MSWAL and AMOS22 support sets, which confirms that sufficient anatomical prompts improve cross-domain generalization. Notably, the Group Prompt module realizes remarkable convergence acceleration and high training efficiency. Using 60 prompts with only 200 epochs, the model reaches competitive Dice values that surpass or approach the 1000-epoch baseline with shorter training duration. Extending training to 1000 epochs further elevates segmentation performance, yet drastically increases runtime. Accordingly, the 60-prompt setup with 200 epochs becomes an efficient and practical solution for clinical applications with limited computing resources. Moreover, for support set construction, the model achieves higher Dice scores on the identical dataset (MSWAL) than on a different dataset (AMOS22).

\subsection{Results of Many-to-one-CSH.}

\begin{figure}
\centering
\includegraphics[width=8cm]{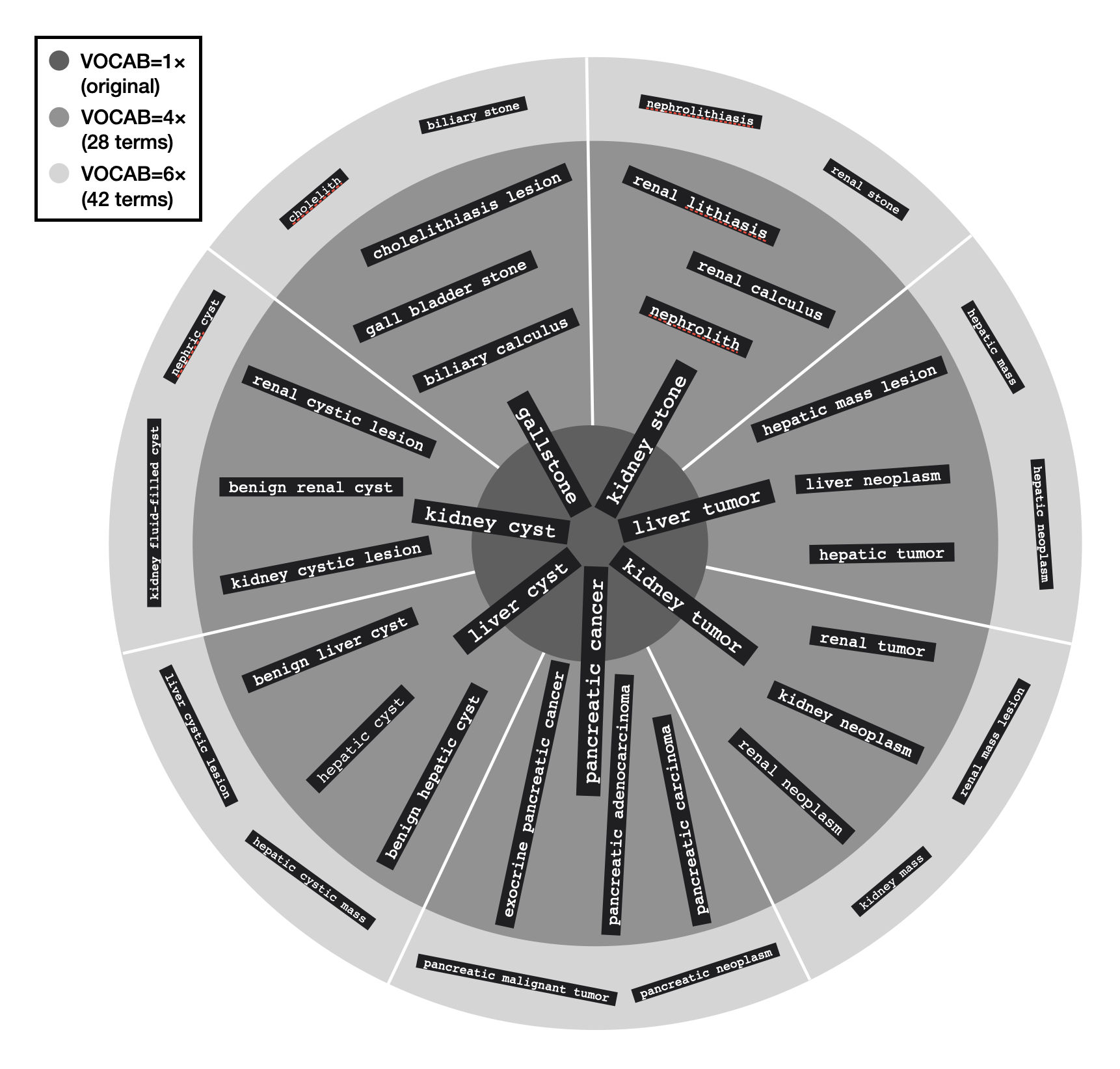}
\caption{Hierarchical expansion of the medical vocabulary registry for the MSWAL dataset. Semantic relevance to the core concept weakens moving outward: the innermost circle is the baseline VOCAB with 7 core medical categories, while the middle and outermost circles include extended clinical and anatomical variants.}
\end{figure}

To evaluate the robustness of our proposed Many-to-One CSH framework against label taxonomy inconsistency, we conducted a rigorous quantitative evaluation under text-prompt perturbations across abdominal lesion and organ datasets. Specifically, during inference on the MSWAL abdominal tumor dataset, the default canonical category names were substituted with unseen clinical synonyms (such as replacing "pancreatic cancer" with "ancreatic malignant neoplasm", which additionally introduces an intentional typographical error). We also excluded the "kidney tumor" class from training to test zero-shot generalization. The baseline vision-language model suffered severe performance degradation under synonym substitution, with zero Dice scores for most tumor categories. This proves conventional models lack tolerance to variations in clinical terminology without dedicated mapping layers. Our CSH method maintained stable segmentation performance under semantic perturbations even without synonym augmentation (\textit{VOCAB}=1×). As the vocabulary mapping dictionary expanded, segmentation accuracy improved significantly, verifying that the many-to-one mapping graph smooths cross-modal latent space. Most notably, on the completely unseen kidney tumor category where the CSH model (\textit{VOCAB}=1×) yields a 0 Dice score, our Many-to-One CSH model equipped with rich synonymous topologies successfully unlocks a zero-shot segmentation ability, achieving a superior Dice score of \textbf{0.139} (\textit{VOCAB}=6×). This improvement indicates that semantic-rich language meta-parameters can learn general visual features for unseen lesions.

\newcommand{\thickhline}{\noalign{\hrule height 1.2pt}} 
\begin{table}[h]
\centering
\setlength{\tabcolsep}{10pt}       
\renewcommand{\arraystretch}{1.2}  
\caption{Ablation study results of Many-to-one-CSH.}\label{tab1}
\resizebox{0.6\textwidth}{!}{%
\begin{tabular}{l|l|cc}
\thickhline 
\multicolumn{2}{c|}{\diagbox{Model}{Dataset}} & MSWAL & FLARE22 \\ \hline
\multicolumn{2}{c|}{SAT (Baseline)}               & 0.014 &  0.841  \\ \hline 
\multirow{3}{*}{CSH} & VOCAB=6$\times$        & \textbf{0.322} & \multirow{3}{*}{0.861} \\
                     & VOCAB=4$\times$        & 0.315 &         \\
                     & VOCAB=1$\times$        & 0.184 &         \\ \hline
\multicolumn{2}{c|}{CLIP-UM}                    & 0.068 & \textbf{0.865}   \\ \thickhline 
\end{tabular}%
}
\end{table}

Furthermore, to verify whether the dynamic CSH adaptation induces negative interference or degradation on standard anatomical structures, we benchmarked the models on the FLARE22 dataset. Overall, both CLIP-UM and our CSH model consistently outperformed the baseline. Specifically, notable performance gains were observed in complex anatomical structures such as Liver (improving from \textbf{0.963} to \textbf{0.972}) and Adrenal Gland (improving from \textbf{0.812} to \textbf{0.835}). This outcome demonstrates that our heavily shared dynamic parameter generation schema successfully retains precise foundational anatomical representations. By maintaining high-precision parsing of standard organs while substantially reinforcing the system's generalization under ambiguous semantics and unseen pathology, this approach provides a highly extensible framework for future universal medical image segmentation.

\section{Conclusion}

In this work, we propose three simple enhancement modules to improve medical lesion segmentation under limited data and computational resources: 1) DT-Morph Combined Loss, which optimizes differentiable topological structures and boundary continuity; 2) 3D Group Prompt, which extracts domain-universal knowledge via curated volumetric image-mask alignment prompts together with text prompts; 3) Many-to-One Conditional Segmentation Head, which brings negligible computational cost and enables fast incremental learning for new categories without sacrificing performance on old tasks. Extensive experiments on multiple abdominal datasets demonstrate its superior cross-domain generalization and robust extensibility to diverse linguistic prompts.

Nevertheless, there are still limitations of current works. Despite outperforming traditional baselines, the model fails to achieve satisfactory accuracy on highly irregular tiny lesions and completely unseen lesions in zero-shot scenarios, limiting its direct deployment in high-risk clinical diagnosis and automatic surgical planning. Given the confidentiality of clinical annotated data and the continuous emergence of new pathological categories with insufficient training samples, we propose a future improvement via progressive label-wise incremental learning. Each category is learned thoroughly in two stages: a few-shot adaptation stage that fine-tunes prompt generators with limited support samples, followed by a full incremental training stage that utilizes all target-class data to stabilize existing anatomical features and adapt to new categories efficiently.

\begin{credits}
\subsubsection{\ackname} We sincerely appreciate the efforts of the authors of SAT [13], UniCDSeg [16], and the CLIP-Driven Universal Model [22] in releasing their source code and pre-trained models to the community, which greatly facilitated the implementation and validation of our method.
\end{credits}
%
%
%
%

\end{document}